\documentclass[aoas,nameyear,seceqn,dvips]{arximspdf}
\usepackage{graphics}
\doi{10.1214/09-AOAS277}
\volume{4}
\issue{1}
\pubyear{2010}
\firstpage{503}
\lastpage{519}

\makeatletter
\newproclaim{remark}{Remark}

  \let\sv@tabnotetext\tabnotetext
  \let\sv@tabnotemark@fmt\tabnotemark@fmt
   \long\def\legend#1{{\let\tabnote@indent\leavevmode\sv@tabnotetext[]{}{#1}}}

\makeatother

\begin{document}
\begin{frontmatter}

\title{Feature selection in omics prediction problems
using cat scores and false nondiscovery rate~control}
\runtitle{LDA feature selection}

\begin{aug}
\author[A]{\fnms{Miika} \snm{Ahdesm\"{a}ki}\ead[label=e1]{miika.ahdesmaki@gmail.com}}
and
\author[B]{\fnms{Korbinian} \snm{Strimmer}\ead[label=e2]{strimmer@uni-leipzig.de}\corref{}}
\runauthor{M. Ahdesm\"{a}ki and K. Strimmer}
\pdfauthor{Miika Ahdesmaki, Korbinian Strimmer}

\affiliation{University of Leipzig, Tampere University of Technology
and University of Leipzig}

\address[A]{Institute for Medical Informatics,\\
\quad Statistics and Epidemiology (IMISE)\\
University of Leipzig\\
H\"{a}rtelstr. 16--18\\
D-04107 Leipzig\\
Germany\\
and\\
Department of Signal Processing\\
Tampere University of Technology\\
P.O. Box 553\\
FI-33101 Tampere\\
Finland\\
\printead{e1}
}
\address[B]{Institute for Medical Informatics,\\
\quad Statistics and Epidemiology (IMISE)\\
University of Leipzig\\
H\"{a}rtelstr. 16--18\\
D-04107 Leipzig\\
Germany\\
\printead{e2}
}

\end{aug}

\received{\smonth{3} \syear{2009}}
\revised{\smonth{7} \syear{2009}}

%
\begin{abstract}
We revisit the problem of feature selection
in linear discriminant analysis (LDA), that is, when features are correlated.
First, we introduce a pooled centroids formulation of the multiclass
LDA predictor function, in which the relative weights of
Mahalanobis-transformed predictors are given
by correlation-adjusted $t$-scores (cat scores).
Second, for feature selection we propose thresholding cat scores
by controlling false nondiscovery rates (FNDR).
Third, training of the classifier is based on
James--Stein shrinkage estimates of correlations and variances,
where regularization parameters are chosen analytically without resampling.
Overall, this results in an effective
and computationally inexpensive
framework for high-dimensional prediction with natural feature selection.
The proposed shrinkage discriminant procedures are implemented in
the R package ``sda'' available from the R repository CRAN.
\end{abstract}

%
\end{frontmatter}

\section{Introduction}

Class prediction of biological samples based on their genetic or
proteomic profile is now a routine task in genomic studies.
Accordingly, many classification methods have been developed to
address the specific statistical challenges presented by these data---see, for example, \citet{SIR08} and \citet{SDB08} for recent reviews.
In particular, the small sample size $n$ renders difficult the training
of the classifier, and the large number of variables $p$ makes it hard
to select suitable features for prediction.

Perhaps surprisingly, despite the many recent innovations in the field of
classification methodology, including the introduction of sophisticated
algorithms for support vector machines and the proposal of ensemble methods
such as random forests,
the conceptually simple approach of linear discriminant analysis (LDA)
and its sibling, diagonal discriminant analysis (DDA),
remain among the most effective procedures also in the domain of
high-dimensional prediction [\citet{Efr08b}; \citet{Hand06}; \citet{Efr75b}].

In order to be applicable for high-dimensional analysis, it has been
recognized early that regularization is essential [\citet{Fri89}].
Specifically, when training the classifier, that is,\ when estimating the
parameters of the discriminant function from training data, particular
care needs to be taken to accurately infer the (inverse) covariance
matrix. A rather radical, yet highly effective
way to regularize covariance estimation in
high dimensions is to set all correlations equal to zero [\citet{BL04}].
Employing a diagonal covariance matrix reduces LDA to the special case of
diagonal discriminant analysis (DDA), also known in the machine learning
community as ``naive Bayes'' classification.

In addition to facilitating high-dimensional estimation of the
prediction function, DDA has one further key
advantage: it is straightforward to conduct feature selection.
In the DDA setting with two classes ($K=2$), it can be shown that
the optimal criterion for
ordering features relevant for prediction are
the $t$-scores between the two group means [e.g., \citet{FF08}], or
in the multiclass setting, the $t$-scores between group means
and the overall centroid.

The nearest shrunken centroids (NSC) algorithm [\citeauthor{THNC02} (\citeyear{THNC02,THNC03})],
commonly known by the name of ``PAM'' after its software implementation,
is a regularized version of DDA with multiclass feature selection.
The fact that PAM has established itself as one of the most popular
methods for classification of gene expression data is ample proof
that DDA-type procedures are indeed very effective for
large-scale prediction problems---see also \citet{BL04} and \citet{Efr08b}.

However, there are now many omics data sets where correlation among
predictors is an essential feature of the data and hence cannot easily
be ignored. For example, this includes proteomics, imaging, and
metabolomics data where correlation among biomarkers is commonplace
and induced
by spatial dependencies and by chemical similarities, respectively.
Furthermore, in many transcriptome measurements
there are correlations among genes within a functional group or pathway
[\citet{AS09}].

Consequently, there have been several suggestions to generalize
PAM to account for correlation. This includes the
SCRDA [\citet{GHT07}], Clanc [\citet{DS07}] and MLDA [\citet{XBP09}] approaches.
All these methods are regularized versions of LDA, and hence offer
automatic provisions for gene-wise correlations.
However, in contrast to PAM and DDA, they lack an efficient and elegant
feature selection scheme, due to problems with multiple
optima in the choice of regularization parameters (SCRDA) and
the large search space for optimal feature subsets (Clanc).

In this paper we present a framework for efficient high-dimensional
LDA analysis.
This is based on three cornerstones. First, we employ
James--Stein shrinkage rules for training
the classifier. All regularization parameters are estimated
from the data in an analytic fashion without resorting to computationally
expensive resampling. Second, we use correlation-adjusted $t$-scores
(cat scores) for feature selection. These scores emerge
from a restructured version of the LDA equations and enable
simple and effective ranking of genes even in the presence of correlation.
Third, we employ false nondiscovery rate thresholding for
selecting features for inclusion in the prediction rule.
As we will show below, this is a highly effective method with similar
performance to the recently proposed ``higher criticism'' approach.

The remainder of the paper is organized as follows. In Sections
\ref{sec2}--\ref{sec5} we detail our framework for shrinkage discriminant
analysis and variable selection. Subsequently, we demonstrate the
effectiveness of our approach by application to a number of
high-dimensional genomic data sets. We conclude with
a discussion and comparison to closely related approaches.


\section{Linear discriminant analysis revisited}\label{sec2}

\subsection{Standard formulation}

LDA starts by assuming a mixture model for the $p$-dimensional data
$\mathbf{x}$,
\[
f(\mathbf{x}) = \sum_{j=1}^K \pi_j f(\mathbf{x}| j) ,
\]
where each of the $K$ classes is represented by a multivariate normal
density
\[
f( \mathbf{x}| k ) = (2 \pi)^{-p/2} | \bolds{\Sigma}|^{-1/2}
 \exp\bigl\{ -\tfrac{1}{2} (\mathbf{x}-\bolds
{\mu}_k)^T
\bolds{\Sigma}^{-1} (\mathbf{x}-\bolds{\mu}_k) \bigr\}
\]
with group-specific centroids $\bolds{\mu}_k$ and a common covariance matrix
$\bolds{\Sigma}$.
The probability of group $k$ given $\mathbf{x}$ is computed from the a priori
mixing weights $\pi_j$ by application of Bayes' theorem,
\[
\operatorname{Pr}(k| \mathbf{x}) = \frac{\pi_k f(\mathbf{x}|
k)}{f(\mathbf{x})} .
\]
We define here the LDA discriminant score as the log posterior
$d_k(\mathbf{x}) = \break\log\{\operatorname{Pr}(k| \mathbf{x})\}$, which
after dropping terms
constant across groups becomes
%
\begin{equation}\label{eq:discrimLDA}
d_k^{\mathrm{LDA}}(\mathbf{x}) =
\bolds{\mu}_k^T \bolds{\Sigma}^{-1} \mathbf{x}-\tfrac{1}{2}
\bolds{\mu}_k^T \bolds{\Sigma}^{-1} \bolds{\mu}_k + \log(\pi_k) .
\end{equation}
Due to the common covariance, $d_k^{\mathrm{LDA}}(\mathbf{x})$ is linear
in $\mathbf{x}$.
Prediction in LDA works by evaluating the discriminant function
at the given test sample $\mathbf{x}$ for all possible~$k$, choosing
the class
maximizing the posterior probability (and hence $d_k^{\mathrm{LDA}}$).

\subsection{Pooled centroid formulation}

We now rewrite the standard form of the LDA predictor function (\ref
{eq:discrimLDA})
with the aim to elucidate the influence of each individual variable in
prediction.
Specifically, we simply add a class-independent constant to the
discriminant function---note that this does not change in any way the prediction.
We compute the pooled mean
\[
\bolds{\mu}_{\mathrm{pool}} = \sum_{j=1}^K \frac{n_j}{n} \bolds{\mu
}_j   ,
\]
representing the overall centroid ($n_j$ is the sample size in group
$j$ and $n = \sum_{j=1}^K n_j$ the total number
of observations) and the corresponding discriminant score
\[
d_{\mathrm{pool}}^{\mathrm{LDA}}(\mathbf{x}) =
\bolds{\mu}_{\mathrm{pool}}^T \bolds{\Sigma}^{-1} \mathbf{x}-\tfrac{1}{2}
\bolds{\mu}_{\mathrm{pool}}^T \bolds{\Sigma}^{-1} \bolds{\mu}_{\mathrm{pool}}  .
\]
The centered score
\[
\Delta_k^{\mathrm{LDA}}(\mathbf{x}) = d_k^{\mathrm{LDA}}(\mathbf{x}) -
d_{\mathrm{pool}}^{\mathrm{LDA}}(\mathbf{x})
\]
can be interpreted as a log posterior ratio and is,
in terms of prediction, completely
equivalent to the original $d_k^{\mathrm{LDA}}(\mathbf{x})$.
After some careful algebra, it simplifies to
%
\begin{equation}\label{eq:multiclasslda}
\Delta_k^{\mathrm{LDA}}(\mathbf{x}) =
\bolds{\omega}_k^T \bolds{\delta}_k(\mathbf{x})
+ \log(\pi_k)
\end{equation}
with feature weight vector
%
\begin{equation}\label{eq:featureweights}
\bolds{\omega}_k = \mathbf{P}^{-1/2} \mathbf{V}^{-1/2} ( \bolds
{\mu}_k - \bolds{\mu}_{\mathrm{pool}} )
\end{equation}
and Mahalanobis-transformed predictors
%
\begin{equation}\label{eq:distfunc}
\bolds{\delta}_k(\mathbf{x}) = \mathbf{P}^{-1/2} \mathbf{V}^{-1/2}
\biggl( \mathbf{x}- \frac{ \bolds{\mu}_k+\bolds{\mu}
_{\mathrm{pool}}}{2}\biggr).
\end{equation}
Here, we have made use of the variance-correlation decomposition of
the covariance matrix $\bolds{\Sigma}= \mathbf{V}^{1/2} \mathbf
{P}\mathbf{V}^{1/2}$,
where $\mathbf{V}= \operatorname{diag}\{\sigma^2_1, \ldots, \sigma^2_p\}$
is a diagonal matrix containing the variances and
$\mathbf{P}= (\rho_{ij})$ is the correlation matrix.

A remarkable property of the above restructuring (\ref
{eq:multiclasslda})--(\ref{eq:distfunc}) of the LDA discriminant
function (\ref{eq:discrimLDA})
is that both $\bolds{\omega}_k$
and $\bolds{\delta}_k(\mathbf{x})$ are \textit{vectors} and not matrices.
Furthermore, note
that $\bolds{\omega}_k$ is not a function of the test data $\mathbf
{x}$ and that its
components control how much each individual variable contributes to the score
$\Delta_k^{\mathrm{LDA}}$ of group~$k$.

\subsection{James--Stein shrinkage rules for learning
the LDA predictor}

In order to train the LDA discriminant function [(\ref{eq:discrimLDA})
or (\ref{eq:multiclasslda})],
we estimate group centroids $\bolds{\mu}_k$ by their empirical means,
and otherwise
rely on three different James--Stein-type shrinkage rules.
Specifically, we employ the following:
\begin{enumerate}
\item for the correlations $\mathbf{P}$ the ridge-type estimator
from \citet{SS05c},
\item for the variances $\mathbf{V}$ the shrinkage estimator from
\citet{OS07a}, and
\item for the proportions $\pi_k$ the frequency estimator
from \citet{HS09}.
\end{enumerate}
All three James--Stein-type estimators are constructed by shrinking
toward suitable targets and analytically minimizing the mean squared error.
The precise formulas are given in Appendix \ref{appA}. For the statistical
background we refer to the above mentioned references.

We remark that the advantages of using James--Stein rules for data analysis
have recently become (again) more appreciated in the literature,
especially in the ``small~$n$, large $p$'' setting,
where James--Stein-type estimators
are very efficient both in a statistical as well as in a computational
sense.
In training of the LDA predictor function by James--Stein shrinkage,
we follow \citet{DS07} and \citet{XBP09}, who give a
comprehensive comparison with competing approaches such as support vector
machines. \citet{SDB08} also implement a shrinkage version of LDA.

\section{Feature selection}

\subsection{A natural variable selection score for LDA}

Following \citet{ZS09}, we define the vector $\bolds{\tau}^{\mathit{adj}}_k$ of
correlation-adjusted $t$-scores
(cat scores) to be a scaled version of the feature weight vector
$\bolds{\omega}_k$:
%
\begin{eqnarray}\label{eq:catscore}
\bolds{\tau}^{\mathit{adj}}_k &\equiv&\biggl(\frac{1}{n_k}-\frac{1}{n}\biggr)^{-1/2}
\bolds{\omega}
_k \nonumber\\
&=& \mathbf{P}^{-1/2} \times\biggl\{\biggl(\frac{1}{n_k}-\frac{1}{n}\biggr) \mathbf
{V}\biggr\}^{-1/2} (
\bolds{\mu}_k - \bolds{\mu}_{\mathrm{pool}} ) \\
&=& \mathbf{P}^{-1/2} \bolds{\tau}_k   .\nonumber
\end{eqnarray}
The vector $\bolds{\tau}_k$ contains the gene-wise gene-specific $t$-scores
between the mean of group $k$ and the pooled mean. Thus, the
correlation-adjusted
$t$-scores ($\bolds{\tau}^{\mathit{adj}}_k$) are decorrelated $t$-scores
($\bolds{\tau}_k$).
If there is no
correlation, $\bolds{\tau}^{\mathit{adj}}_k$ reduces to $\bolds{\tau}_k$.
The factor $(\frac{1}{n_k}-\frac{1}{n})^{-1/2}$ in equation (\ref{eq:catscore})
standardizes the error of $\hat{\bolds{\mu}}_k- \hat{\bolds{\mu}}_{\mathrm{pool}}$, and
is the same as in PAM [\citet{THNC03}].
Note the minus
sign, which is due to correlation $\sqrt{n_k/n}$ between $\hat{\bolds{\mu}}
_k$ and
$\hat{\bolds{\mu}}_{\mathrm{pool}}$.\footnote{The
plus sign in the original PAM paper
[\citet{THNC02}, page 6567] is a typographic error.}

In DDA approaches, such as PAM, regularized estimates of the $t$-scores
$\bolds{\tau}_k$ are employed
for feature selection.
From equations~(\ref{eq:multiclasslda})--(\ref{eq:distfunc}) it follows directly
that the cat scores $\bolds{\tau}^{\mathit{adj}}_k$ provide the most natural
generalization
in the LDA setting (see also Remark \ref{remA}).

As a summary score to measure the total impact of feature
$i \in\{1, \dots, p\}$, we
use
%
\begin{equation}
S_i = \sum_{j=1}^K (\tau^{\mathit{adj}}_{i,j})^2   ,
\label{eq:summaryscore}
\end{equation}
that is,\ the squared $i$th component of the cat score vector
$\bolds{\tau}^{\mathit{adj}}_k = (\tau^{\mathit{adj}}_{1,k}, \dots, \tau^{\mathit{adj}}_{p,k})^T$
summed across the $K$ groups.
For comparison, PAM uses the criterion
%
\begin{equation}
\label{eq:pamscore}
S_i^{'} = \max_{j=1, \dots, K}(|\tau_{i,j}|)   .
\end{equation}
Using the squared sum of the group-specific cat scores
in $S_i$ rather than taking the maximum over the absolute
values as in $S_i^{'}$ has two distinct advantages. First,
the sample distribution of the estimated $S_i$ is more tractable,
being approximately $\chi^2$. Second, if a feature is discriminative
with regard to more than one group, this additional information
is not disregarded.

\subsection{Feature selection by controlling the false nondiscovery rate}

When constructing an efficient classifier, it is desirable to
eliminate
features that provide no useful information for discriminating among
classes. The conventional but computationally tedious
approach is to choose the optimal threshold by estimating
the prediction error via cross-validation along a grid of possible
threshold values. Faster alternative thresholding procedures include
``higher criticism'' [\citet{DJ08}], ``FAIR'' [\citet{FF08}]
and ``Ebay'' [\citet{Efr08a}]. The latter two methods are primarily
developed with the correlation-free setting and $t$-scores in mind
(however, ``Ebay'' also offers correlation corrections for prediction errors).

Here, we advocate using the false discovery rate (FDR) framework
to select features for classification.
We emphasize, however, that in the problem of constructing classifiers
the FDR approach can\textit{not} be applied in the same fashion as
in differential expression. In the latter case,
the aim is to compile a set of genes one has confidence in to be differentially
expressed. This is controlled by the FDR criterion. In contrast, when furnishing
classifiers, one aims at identifying with confidence the set of null features
that are not informative with regard to group separation, in order to eliminate
them from the classifier. This is controlled by the false \textit{non}discovery
rate, FNDR. For a discussion of the relation between FDR and FNDR see,
for example,
\citet{Str08c}.

\begin{figure}

\includegraphics{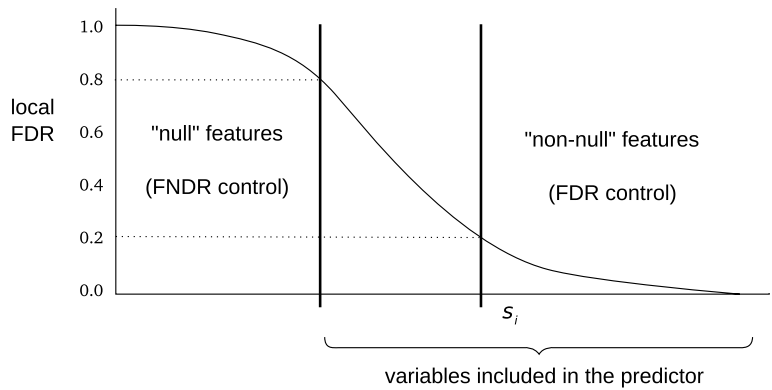}

\caption{Local false discovery rates as a function of the
summary score $S_i$. There are three distinct areas:
an acceptance and a rejection zone, which are separated by a ``buffer
zone'' in
the middle. Note that the features to be included in the classifier by
FNDR control of the null genes
form a superset of the differentially expressed genes controlled by FDR.}
\label{fig:fdr.curve}
\end{figure}

This subtle but important distinction is best illustrated by referring
to Figure~\ref{fig:fdr.curve}, which plots the local FDR $\mathrm{fdr}(S_i) = \operatorname{Prob}(\mbox{``null''} | S_i)$ computed for (and from)
the statistic $S_i$ of feature $i$.
In a list of differentially
expressed genes we decide to include, say, genes $i$ with $\mathrm{fdr}(S_i) < 0.2$.
A similar constraint on the local false nondiscovery rate, $\mathrm{fndr}(S_i) < 0.2$,
gives a confidence set of the null genes. The local false discovery
and local false nondiscovery rates add up to one,
$\mathrm{fndr}(S_i) = 1- \mathrm{fdr}(S_i)$.
Hence, the set of features to be retained in the classifier have
local false discovery rates smaller than 0.8---instead of 0.2.
Thus, the features included in the predictor form a superset of the
differentially expressed variables.
A similar argument applies when using distribution-based Fdr
($q$-values) and
Fndr values.

\begin{table}[b]
\tablewidth=230pt
\caption{The general feature selection score $S_i$ and special cases thereof}\label{tab:fselscore}
\begin{tabular*}{230pt}{@{\extracolsep{\fill}}lcc@{}}
\hline
$\bolds{S_i =} $ & $\bolds{K}$ \textbf{arbitrary} & $\bolds{K=2}$ \\
\hline
Correlation present: & $\sum_{j=1}^K (\tau^{\mathit{adj}}_{i,j})^2$ & $2 (\tau
^{\mathit{adj}}_i)^2$ \\[5pt]
No correlation ($\mathbf{P}= \mathbf{I}$): & $\sum_{j=1}^K \tau
_{i,j}^2$ & $2
\tau_i^2$ \\
\hline
\end{tabular*}
\end{table}

In short, our proposal is to identify the null features by
controlling (local) FNDR, and subsequently using all features except the
identified null set in prediction. For estimating FDR quantities, we use
the semiparametric approach outlined in \citet{Str08c}. Note that this
and other FDR procedures assume that there are enough null features
so that the null model can be properly estimated [\citet{Efr04}].

\section{Special cases}

\subsection{Two groups}

For $K=2$ the cat score $\bolds{\tau}^{\mathit{adj}}_k$ between the group
centroid and the pooled mean
reduces to the cat score between the two means:
\begin{eqnarray*}
\bolds{\tau}^{\mathit{adj}}_1 &=& \mathbf{P}^{-1/2} \times\biggl\{\biggl(\frac
{1}{n_1}-\frac
{1}{n_1+n_2}\biggr) \mathbf{V}\biggr\}^{-1/2}
\biggl( \bolds{\mu}_1 - \biggl(\frac{n_1}{n} \bolds{\mu}_1 +\frac{n_2}{n}
\bolds{\mu}_2\biggr) \biggr) \\
&=& \mathbf{P}^{-1/2} \times\biggl\{\biggl(\frac{1}{n_1}+\frac{1}{n_2}\biggr)\mathbf
{V}\biggr\}^{-1/2} (
\bolds{\mu}_1 -\bolds{\mu}_2) .
\end{eqnarray*}
Note that $\bolds{\tau}^{\mathit{adj}}_1 = - \bolds{\tau}^{\mathit{adj}}_2$.
The feature selection score $S_i$ reduces to the squared cat score between
the two means; cf. Table~\ref{tab:fselscore}.
Likewise, for $K=2$ the difference
$\Delta_1^{\mathrm{LDA}}(\mathbf{x})-\Delta_2^{\mathrm{LDA}}(\mathbf{x})$
reduces to
$\bolds{\omega}^T \bolds{\delta}(\mathbf{x}) + \log(\frac{\pi
_1}{\pi_2})$ with
$\bolds{\omega}= \mathbf{P}^{-1/2} \mathbf{V}^{-1/2} ( \bolds{\mu
}_1 - \bolds{\mu}_2 )$
and $\bolds{\delta}(\mathbf{x}) = \mathbf{P}^{-1/2} \mathbf
{V}^{-1/2} ( \mathbf{x}- \frac{ \bolds{\mu}
_1+\bolds{\mu}_2}{2})$.

For extensive discussion of the two group case, including comparison of
gene rankings with many other test statistics, we refer to \citet{ZS09}.

\subsection{Vanishing correlation}

In case of no correlation ($\mathbf{P}= \mathbf{I}$),
the cat scores reduce (by construction) to
standard $t$-scores between the two centroids of interest,
either between the group and the pooled mean (general $K$)
or between the two groups ($K=2$).
The gene summary $S_i$ reduces to the sum of the respective squared
$t$-scores (Table~\ref{tab:fselscore}).
The discriminant function reduces to the standard form of
diagonal discriminant analysis. The pooled centroids formulation of
LDA reduces to that of PAM (except for the shrinkage of the
means present in PAM but not in our approach).

\section{Remarks}\label{sec5}

\renewcommand{\theremark}{\Alph{remark}}

\begin{remark}[(Definition of feature weights)]\label{remA}
The definition of feature weights according to equation (\ref{eq:featureweights})
is most natural. Other ways of splitting up the product
$\bolds{\omega}_k^T \bolds{\delta}_k(\mathbf{x})$ lead to various
inconsistencies.
For example, instead of using $\bolds{\omega}_k = \bolds{\Sigma
}^{-1/2} ( \bolds{\mu}_k -
\bolds{\mu}_{\mathrm{pool}} )$,
it has been suggested to consider $\bolds{\Sigma}^{-1} ( \bolds{\mu
}_k - \bolds{\mu}
_{\mathrm{pool}} )$,
for example, in \citet{WT09}, page 627. However, this choice implies
that for
the case of no correlation variable selection would be based on
$\mathbf{V}^{-1} ( \bolds{\mu}_k - \bolds{\mu}_{\mathrm{pool}} )$
rather than on $t$-scores.

Furthermore, dividing the inverse correlation $\mathbf{P}^{-1}$ equally
between equations (\ref{eq:featureweights}) and (\ref{eq:distfunc}) greatly
simplifies interpretation: feature selection takes place on the
level of centered, scaled as well as decorrelated predictors $\bolds
{\delta}
_k(\mathbf{x})$.
Note that this interpretation is not hampered by the fact that the decorrelation
involves all features, because typically there is no substantial
correlation between
nonnull and null features, so that the overall correlation matrix decomposes
into correlation within null and within nonnull variables.
\end{remark}

\begin{remark}[(Grouping of features)]\label{remB}
Using cat scores for feature selection also greatly facilitates
the grouping of features. Specifically, adding the squared
cat scores of each feature contained in a given set (e.g., gene sets specified
by biochemical pathways) leads to
Hotelling's $T^2$; see \citet{ZS09}.
Note that if another decomposition than that of
equation (\ref{eq:featureweights}) and (\ref{eq:distfunc})
was used, the connection of cat scores with Hotelling's $T^2$
would be lost.
\end{remark}

\begin{remark}[(FDR methods for feature selection)]\label{remC}
The usefulness of false discovery rates for feature selection in
prediction is disputed, for example, in \citet{DJ08}. What we show
here is
that the unfavorable performance is due to naive application of FDR,
leading to the elimination of too many predictors. If instead FNDR is
controlled to determine the null-features to be excluded from the
discriminant function, then much more efficient prediction rules are obtained.
\end{remark}

\begin{remark}[(Fast computation of matrix square root)]\label{remD}
The inverse square root of the
correlation matrix, required in equations~(\ref{eq:discrimLDA}) and
(\ref{eq:catscore}), can be computed very efficiently for the
James--Stein shrinkage estimator; see \citet{ZS09} for details.
\end{remark}

\begin{remark}[(Normalizing the null model)]\label{remE}
Estimating false discovery rates using summary
scores $S_i$ (\ref{eq:summaryscore}) assumes as null model a $\chi
^2$-distribution with
unknown parameters. To employ standard FDR software,
we apply the cube-root transformation,
which provides a normalizing transform for the $\chi^2$-distribution
[\citet{WH31}].
\end{remark}

\section{Results}

We now illustrate our shrinkage DDA and LDA approaches with
variable selection using cat scores and FNDR control
by analyzing a number of reference data examples, and compare
our results with that of competing approaches. We also investigate the
performance of FNDR feature selection in comparison with that of
``higher criticism''
[\citet{DJ08}].

\subsection{\texorpdfstring{Singh et al. (\protect\citeyear{SF02}) gene expression
data}{Singh et al. (2002) gene expression data}}

\begin{figure}

\includegraphics{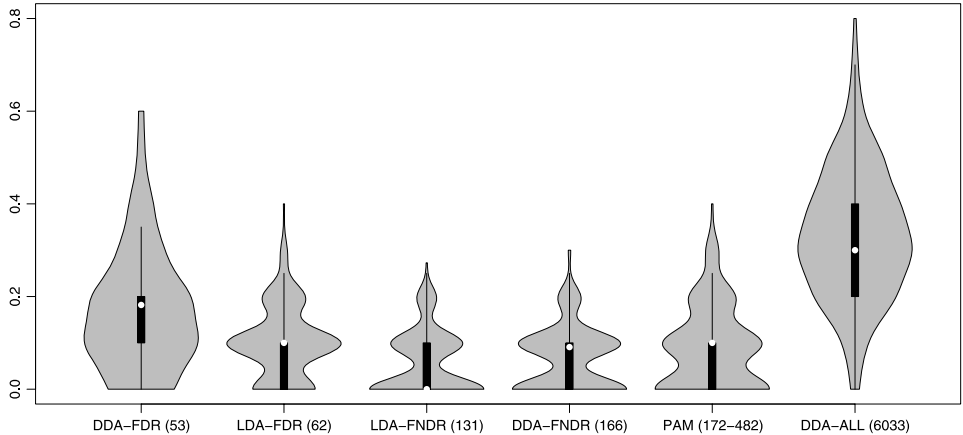}

\caption{Violin plots of prediction error rates of various
classification methods
for the Singh \textit{et al.} (\protect\citeyear{SF02}) data. The violin plot is a generalization
of the box plot, showing the median and
upper and lower quartiles, as well as the density. Underlying each plot
are 200 estimates of prediction error computed from the 200 splits arising
from balanced 10-fold cross-validation with 20 repetitions.
The number in round brackets
indicates the number of selected features. See also Table~\protect\ref
{tab:singh.results}.}
\label{fig:singh.violin}
\end{figure}

First, we investigated the prostate cancer data set of
\citet{SF02}. This consists of gene expression measurements
of $p=6033$ genes for $n=102$ patients, of which 52 are cancer
patients and 50 are healthy (thus, $K=2$). To facilitate
cross-comparison, we analyzed
the data exactly in form as used in \citet{Efr08b}.
Our results are summarized in Table~\ref{tab:singh.results},
and corresponding violin plots [\citet{HN98}] are shown in Figure~\ref{fig:singh.violin}.

\begin{table}[b]
\tablewidth=220pt
\caption{Prediction errors and number of selected features for Singh
\textit{et al.} (\protect\citeyear{SF02}) gene expression data.
The number in the round brackets is the estimated standard error}
\label{tab:singh.results}
\begin{tabular*}{220pt}{@{\extracolsep{\fill}}lcc@{}}
\hline
\textbf{Method} & \textbf{Prediction Error} & \textbf{Features} \\
\hline
Ebay & 0.092\phantom{0 (0.0000)} & \phantom{000--0}51 \\
DDA-FDR & 0.1682 (0.0093) & \phantom{000--0}53 \\
LDA-FDR & 0.0989 (0.0056) & \phantom{000--0}62 \\
LDA-FNDR & \textbf{0.0550} (0.0048) & \phantom{000--}131 \\
DDA-FNDR & 0.0640 (0.0049) & \phantom{000--}166\\
PAM & 0.0859 (0.0063) & 172--482 \\
DDA-ALL & 0.3327 (0.0099) & \phantom{00--}6033 \\
\hline
\end{tabular*}
\legend{The prediction error of Ebay is taken from \citet{Efr08b}.}
\end{table}

Initially, we assumed zero correlation and applied the shrinkage DDA method.
By controlling the local FNDR to be smaller or equal
than 0.2, we determined
that 5867 genes were null genes, hence, that 166 genes needed to be
included in the prediction rule. For comparison, a local FDR cutoff
on the same level
yielded only 53 genes, lacking the 103 genes in the ``buffer zone''
between the two cutoffs (cf. Figure~\ref{fig:fdr.curve}). Note that we
recommend using
the larger FNDR-based feature set, not just the 53 genes considered to be
differentially expressed.

We estimated the prediction error of the resulting classification rule
using balanced 10-fold cross-validation with 20 repetitions.
For each of the in total 200 splits
we trained a new prediction rule and estimated new feature rankings
and FDR statistics, thereby including the selection process in the error
estimate to avoid overoptimistic results [\citet{AL2002}].
The number of selected features shown in Table~\ref{tab:singh.results}
is based on the complete data. However, for estimation of prediction error
for each of the splits a new set of features was determined.

For the FNDR-based cutoff with 166 included features,
we obtained an estimate of the prediction error of 0.0640,
whereas for the naive FDR cutoff resulting in 53 predictors, the error is
much higher (0.1682).
For comparison, we also computed the error using all 6033
features, yielding a massive 0.3327.
The PAM program selected between 172 and 482 genes for inclusion in its
predictor
with error rate 0.0859 (note that the number of selected features
by the PAM algorithm is highly variable and differs from run to
run even for the same data set).
According to
\citet{Efr08b}, the Ebay approach used 51 genes for prediction
with error rate 0.092.

If correlation was taken into account, that is, if the order of ranking
was determined
by cat rather than $t$-scores, interestingly both the number of differentially
expressed genes and of the null genes increases, implying that the ``buffer
zone'' shown in Figure~\ref{fig:fdr.curve} becomes smaller.
Thus, the LDA classifier with FNDR cutoff contained for this data fewer
predictors
(131) but at the same
time nevertheless achieved the smallest overall prediction error
(Figure~\ref{fig:singh.violin}).

\subsection{Performance for multiclass reference data sets}

\begin{table}
\tablewidth=318pt
\caption{Estimated prediction errors for several multiclass reference
data sets}\label{tab:otherdata.results}
\begin{tabular*}{318pt}{@{\extracolsep{\fill}}lcccc@{}}
\hline
\textbf{Data} & \textbf{Method} & \textbf{Prediction error} & \textbf{Features} & \textbf{DE} \\
\hline
Lymphoma & DDA-FNDR & 0.0517 (0.0062) & \phantom{000--00}162 & \phantom{0}0 \\
($K=3$, $n=62$, & LDA-FNDR & \textbf{0.0036} (0.0018) & \phantom{000--00}392 & 55 \\
\quad$p=4026$) & PAM & 0.0254 (0.0045) & 2796--3201 & \\[5pt]
SRBCT & DDA-FNDR & 0.0007 (0.0007) & \phantom{0000--00}90 & 62 \\
($K=4$, $n=63$, & LDA-FNDR & \textbf{0.0000} (0.0000) & \phantom{0000--00}89 & 76 \\
\quad$p=2308$) & PAM & 0.0145 (0.0034) & \phantom{0000}39--87 & \\[5pt]
Brain & DDA-FNDR & 0.1892 (0.0146) & \phantom{000--000}33 & \phantom{0}8 \\
($K=5$, $n=42$, & LDA-FNDR & \textbf{0.1525} (0.0120) & \phantom{0000--0}102 & 23 \\
\quad$p=5597$) & PAM & 0.1939 (0.0112) & \phantom{0}197--5597 & \\
\hline
\end{tabular*}
\legend{The last column (DE) shows the number of differentially expressed genes,
which equals the number of significant features
if FDR rather than FNDR is used as criterion.}
\end{table}

For extended comparison we applied our approach to a number of
further reference data sets. In particular, we analyzed
gene expression data for lymphoma [\citet{AED00}], small round blue
cell tumors
(SRBCT) [\citet{KW01}] and brain cancer [\citet{PTG02}]. The data
sets have
in common that all contain more than two classes, thus allowing to
study
the multi-class summary statistic (\ref{eq:summaryscore}).
A summary of the results obtained by shrinkage LDA/DDA
and FNDR feature selection and by PAM
is given in Table~\ref{tab:otherdata.results}.

The \citet{KW01} data are very easy to classify. All methods
performed equally well on this data, with no substantial
difference between the LDA and DDA approaches.

For the lymphoma data set the PAM approach failed to identify a
compact set of predictive features. In contrast, the FNDR
approach selects a comparatively small number of genes
both in the LDA and DDA case. Intriguingly, for this data
there were no differentially expressed genes, if correlation is
ignored, yet the FNDR criterion yielded 162 nonnull features.

The brain data set is the largest and most difficult data set.
Again, the PAM approach failed to determine a stable set of
features, whereas FNDR control yielded a compact set of informative
predictors. Here, as well as for the lymphoma data, the LDA approach
clearly outperformed the DDA approaches in terms of prediction error.

\subsection{Comparison with \textup{``}higher criticism\textup{''} feature selection}

Using the data examples above, we demonstrated that
feature selection based
on simple FDR cutoffs is not sufficient for prediction.
In particular, if features are weak and sparse, it may
easily happen that no predictor has sufficiently small false
discovery rate to be called significant (cf. the lymphoma data).

\begin{table}
\tablewidth=305pt
\caption{Estimated prediction errors employing higher criticism as
feature selection criterion}\label{tab:hc.results}
\begin{tabular*}{305pt}{@{\extracolsep{\fill}}lcccc@{}}
\hline
\textbf{Data} & \textbf{Method} & \textbf{Prediction error} & \textbf{Features} & \textbf{local FDR} \\
\hline
Prostate & DDA-HC & 0.0707 (0.0055) & 129 & 0.69 \\
& LDA-HC & 0.0497 (0.0045) & 116 & 0.73 \\
Lymphoma & DDA-HC & 0.0185 (0.0038) & 179 & 1.00 \\
& LDA-HC & 0.0000 (0.0000) & 345 & 0.78 \\
SRBCT & DDA-HC & 0.0035 (0.0016) & 138 & 1.00 \\
& LDA-HC & 0.0007 (0.0007) & 174 & 1.00 \\
Brain & DDA-HC & 0.1572 (0.0118) & \phantom{0}33 & 0.77 \\
& LDA-HC & 0.1417 (0.0108) & 131 & 1.00 \\
\hline
\end{tabular*}
\legend{The last column (local FDR) shows the local FDR of the least
significant feature.}
\end{table}

In such a setting \citet{DJ08} suggest as alternative to FDR-based thresholding
the ``higher criticism'' (HC) approach. The HC criterion is based on
$p$-values. For each feature, the $p$-value is centered and
standardized using the estimated mean and variance of the corresponding
order statistic. The optimal threshold is determined
as the maximum of the absolute HC scores within a fraction (say, 10\%) of
the top ranking features [\citet{DJ08}].

Our feature selection approach based on FNDR control shares with HC
that we aim to overcome the limitations resulting from naive application
of FDR-based feature selection. For this reason, it is instructive
to investigate our
shrinkage prediction rule in combination with the HC thresholding procedure.
The $p$-values underlying the HC objective function were obtained by fitting
a two-component mixture model, so that the same empirical null model
was used as in the FDR analysis.

The results are given in Table~\ref{tab:hc.results}.
Again, in all cases the LDA approach
using cat scores for feature selection leads to smaller prediction
error than employing DDA and $t$-scores.
Remarkably, the performance of the FNDR and HC approach are on an equal level,
implying that efficient feature selection is indeed possible
\textit{within} the FDR framework. The set of features selected by
HC is, on average, a bit smaller than that chosen by FNDR, and larger than
the FDR-based set, which indicates that the HC threshold is typically
situated in the ``buffer zone'' of Figure~\ref{fig:fdr.curve}.

\section{Discussion}

\subsection{Shrinkage discriminant analysis and feature selection}

In this paper we have revisited high-dimensional shrinkage
discriminant analysis
and presented a very efficient procedure for prediction.
Our approach contains three distinct elements:
\begin{itemize}
\item the use of James--Stein shrinkage for training the
predictor,
\item feature ranking based on cat scores, and
\item feature selection based on FNDR thresholding.
\end{itemize}
Employing James--Stein shrinkage estimators is efficient both
from a statistical as well as from a computational perspective.
Note that shrinkage is used here only as a means to improve
the estimated parameters, but not for model selection as in
the approaches by \citet{THNC02} and \citet{GHT07}.

The correlation-adjusted $t$-score (cat score) emerges as a natural
gene ranking criterion in the presence of correlation among
predictors [\citet{ZS09}]. Here we have shown how to employ cat scores
in the
multi-class LDA setting and demonstrated on high-dimensional data that
using cat scores rather than $t$-scores leads to a more effective choice
of predictors. We note that the order of ranking induced by the cat and
$t$-scores, respectively,
may differ substantially. Hence, univariate thresholding procedures to select
interesting features will differ, even if the testing procedures
account for
dependencies.

Finally, we propose feature selection by
controlling FNDR rather than FDR and show that this is as efficient
in terms of predictive accuracy as when ``higher criticism'' is employed.
Moreover, we explain why variable selection
based on FDR leads to inferior prediction rules.

\subsection{Recommendations}

For extremely high-dimensional data, estimating correlation is very
difficult, hence, in this instance we recommend to conduct diagonal
discriminant analysis [see also \citet{BL04}]. From our analysis it
is clear the shrinkage DDA as proposed here, combined with variable
selection by control of FNDR or HC, is most effective. In contrast
to the PAM approach, no randomization procedures are involved and, hence,
the prediction rule and the number of selected features are stable.

In all other cases we recommend a full shrinkage LDA analysis,
with feature selection based on cat scores. While this approach
is computationally more expensive than the
shrinkage DDA approach, it has a significant impact
on predictive accuracy. Typically, in comparison with DDA, taking account
of correlation either leads to more compact feature sets
or improved prediction error, or both.
Furthermore, relative to competing full LDA
approaches, such as \citet{GHT07}, our procedure is
computationally fast, due to the avoidance of computer-intensive
procedures such as resampling.

\begin{appendix}

\section{James--Stein shrinkage estimators for
training the LDA predictor}\label{appA}

For ``small $n$, large $p$'' inference of the LDA predictor function
(\ref{eq:discrimLDA}) and (\ref{eq:multiclasslda})
and the cat score (\ref{eq:catscore})
we rely on three different James--Stein-type estimators.

The correlation matrix is estimated by shrinking
empirical correlations $r_{ij}$ toward zero,
\[
r_{ij}^{\mathrm{shrink}} = (1-\hat\lambda_1) r_{ij}   ,
\]
with estimated intensity
\[
\hat\lambda_1 = \min\biggl(1,
\frac{\sum_{i\ne j}\widehat{\operatorname{Var}}(r_{ij}) }{\sum_{i\ne j}
r_{ij}^2} \biggr)
\]
[\citet{SS05c}].

The variances are estimated by shrinking the empirical estimates $v_i$
toward their median,
\[
v_{i}^{\mathrm{shrink}} = \hat\lambda_2 v_{\mathrm{median}} + (1-\hat
\lambda_2) v_i   ,
\]
using
\[
\hat\lambda_2 = \min\biggl(1,
\frac{\sum_{i=1}^p \widehat{\operatorname{Var}}(v_{i}) }{\sum_{i=1}^p
(v_{i} - v_{\mathrm{median}})^2}\biggr)
\]
[\citet{OS07a}].

The class frequencies are estimated following \citet{HS09}
by
\[
\hat\pi_j^{\mathrm{shrink}} = \hat\lambda_3 \frac{1}{K} + (1-\hat
\lambda_3) \frac{n_j}{n}   ,
\]
using
\[
\hat\lambda_3 =
\frac{ 1- \sum^K_{j=1} (n_j/n)^2}{(n-1)
\sum^K_{j=1} (1/K - n_j/n)^2}   .
\]

\section{Relationship to other DDA and LDA approaches}

Our proposed shrinkage discriminant approach is closely linked
to a number of recently proposed methods.

\subsection*{NSC}
The NSC/PAM classification rule was first presented in \citet{THNC02}
and later discussed in more statistical detail in \citet{THNC03}.
PAM is a DDA approach, so no gene-wise correlations are taken into account.
Genes are ranked according to equation (\ref{eq:pamscore}), and
feature selection is determined by soft-thresholding, using
prediction error estimated by cross-validation as optimality criterion.

\subsection*{Ebay}
The ``Ebay'' approach of \citet{Efr08b} is also a DDA approach.
Feature selection is based on an empirical Bayes model that links
prediction error with false discovery rates. Thus, it is very
similar to PAM but computationally and statistically more efficient.
In addition, the ``Ebay'' algorithm provides correlation corrections of
prediction errors; see Section 5 in \citet{Efr08a}.

\subsection*{Clanc and MLDA}
The ``Clanc'' algorithm is described in \citet{DS07} and
the ``modified LDA'' (MLDA) in \citet{XBP09}. Both methods are based
on the LDA framework, and both use James--Stein shrinkage to
estimate the pooled covariance matrix. MLDA uses standard $t$-scores
for feature selection, whereas Clanc employs a greedy algorithm
search to find optimal subsets of features based on a multivariate
criterion.

\subsection*{SCRDA}
The ``shrunken centroids regularized discriminant analysis''\break
(SCRDA)
procedure is described in \citet{GHT07} and uses a similar
soft-thresholding procedure for variable selection as PAM.
The covariance matrix is estimated by a ridge estimator.
Regularization and feature selection parameters are simultaneously determined
by cross-validation. The main issues with SCRDA are the computational
expense and problems in finding unique parameters [\citet{GHT07}].

\section{Computer implementation}

We have implemented the proposed shrinkage discriminant procedures
(both DDA and LDA) and the associated
FNDR and higher criticism variable selection
in the R package ``sda,'' which
is freely available under the terms of the GNU General Public License
(version 3 or later) from CRAN
(\url{http://cran.r-project.org/web/packages/sda/}).
\end{appendix}

\section*{Acknowledgments}
We thank Verena Zuber for critical comments and
helpful discussion. M. A. is grateful to the Alexander von Humboldt Foundation
for a postdoctoral research fellowship.

\printaddresses


\begin{thebibliography}{99}

\bibitem[\protect\citeauthoryear{Ackermann and Strimmer}{2009}]{AS09}
\textsc{Ackermann, M.} and \textsc{Strimmer, K.} (2009).
A general modular framework for gene set enrichment.
\textit{BMC Bioinformatics} \textbf{10} 47.

\bibitem[\protect\citeauthoryear{Alizadeh et~al.}{2000}]{AED00}
\textsc{Alizadeh, A.~A., Eisen, M.~B., Davis, R.~E., Ma, C., Lossos,
I.~S., Rosenwald,
A., Boldrick, J.~C., Sabet, H., Tran, T., Yu, X., Powell, J.~I., Yang, L.,
Marti, G.~E., Moore, T., Hudson, J., Lu, L., Lewis, D.~B., Tibshirani, R.,
Sherlock, G., Chan, W.~C., Greiner, T.~C., Weisenburger, D.~D., Armitage,
J.~O., Warnke, R., Levy, R., Wilson, W., Grever, M.~R., Byrd, J.~C.,
Botstein, D., Brown, P.~O.} and \textsc{Staudt, L.~M.} (2000).
Distinct types of diffuse large {B}-cell lymphoma identified by gene
expression profiling.
\textit{Nature} \textbf{403} 503--511.

\bibitem[\protect\citeauthoryear{Ambroise and {McLachlan}}{2002}]{AL2002}
\textsc{Ambroise, C.} and \textsc{{McLachlan}, G.~J.} (2002).
Selection bias in gene extraction on the basis of microarray
gene-expression data.
\textit{Proc. Natl. Acad. Sci. USA} \textbf{99} 6562--6566.

\bibitem[\protect\citeauthoryear{Bickel and Levina}{2004}]{BL04}
\textsc{Bickel, P.~J.} and \textsc{Levina, E.} (2004).
Some theory for {Fisher's} linear discriminant function, `naive
{Bayes},' and some alternatives when there are many more variables than
observations.
\textit{Bernoulli} \textbf{10} 989--1010.
\MR{2108040}

\bibitem[\protect\citeauthoryear{Dabney and Storey}{2007}]{DS07}
\textsc{Dabney, A.~R.} and \textsc{Storey, J.~D.} (2007).
Optimality driven nearest centroid classification from genomic data.
\textit{PLoS ONE} \textbf{2} e1002.

\bibitem[\protect\citeauthoryear{Donoho and Jin}{2008}]{DJ08}
\textsc{Donoho, D.} and \textsc{Jin, J.} (2008).
Higher criticism thresholding: Optimal feature selection when useful
features are rare and weak.
\textit{Proc. Natl. Acad. Sci. USA} \textbf{105} 14790--15795.

\bibitem[\protect\citeauthoryear{Efron}{1975}]{Efr75b}
\textsc{Efron, B.} (1975).
The efficiency of logistic regression compared to normal discriminant
analysis.
\textit{J. Amer. Statist. Assoc.} \textbf{70} 892--896.
\MR{0391403}

\bibitem[\protect\citeauthoryear{Efron}{2004}]{Efr04}
\textsc{Efron, B.} (2004).
Large-scale simultaneous hypothesis testing: The choice of a null
hypothesis.
\textit{J.~Amer. Statist. Assoc.} \textbf{99} 96--104.
\MR{2054289}

\bibitem[\protect\citeauthoryear{Efron}{2008a}]{Efr08b}
\textsc{Efron, B.} (2008a).
Empirical {Bayes} estimates for large-scale prediction problems.
Technical report, Dept. Statistics, Stanford Univ.

\bibitem[\protect\citeauthoryear{Efron}{2008b}]{Efr08a}
\textsc{Efron, B.} (2008b).
Microarrays, empirical {Bayes}, and the two-groups model.
\textit{Statist. Sci.} \textbf{23} 1--22.
\MR{2431866}

\bibitem[\protect\citeauthoryear{Fan and Fan}{2008}]{FF08}
\textsc{Fan, J.} and \textsc{Fan, Y.} (2008).
High-dimensional classification using features annealed independence
rules.
\textit{Ann. Statist.} \textbf{36} 2605--2637.
\MR{2485009}

\bibitem[\protect\citeauthoryear{Friedman}{1989}]{Fri89}
\textsc{Friedman, J.~H.} (1989).
Regularized discriminant analysis.
\textit{J. Amer. Statist. Assoc.} \textbf{84} 165--175.
\MR{0999675}

\bibitem[\protect\citeauthoryear{Guo, Hastie and Tibshirani}{2007}]{GHT07}
\textsc{Guo, Y., Hastie, T.} and \textsc{Tibshirani, T.} (2007).
Regularized discriminant analysis and its application in microarrays.
\textit{Biostatistics} \textbf{8} 86--100.

\bibitem[\protect\citeauthoryear{Hand}{2006}]{Hand06}
\textsc{Hand, D.~J.} (2006).
Classifier technology and the illusion of progress.
\textit{Statist. Sci.} \textbf{21} 1--14.
\MR{2275965}

\bibitem[\protect\citeauthoryear{Hausser and Strimmer}{2009}]{HS09}
\textsc{Hausser, J.} and \textsc{Strimmer, K.} (2009).
Entropy inference and the James--Stein estimator, with application
to nonlinear gene association networks.
\textit{J. Mach. Learn. Res.} \textbf{10} 1469--1484.

\bibitem[\protect\citeauthoryear{Hintze and Nelson}{1998}]{HN98}
\textsc{Hintze, J.~L.} and \textsc{Nelson, R.~D.} (1998).
Violin plots: A box plot-density trace synergism.
\textit{Amer. Statist.} \textbf{52} 181--184.

\bibitem[\protect\citeauthoryear{Khan et~al.}{2001}]{KW01}
\textsc{Khan, J., Wei, J.~S., Ringner, M., Saal, L.~H., Ladanyi, M.,
Westermann, F.,
Berthold, F., Schwab, M., Antonescu, C.~R., Peterson, C.} and \textsc
{Meltzer, P.~S.}
(2001).
Classification and diagnostic prediction of cancers using gene
expression profiling and artificial neural networks.
\textit{Nature Med.} \textbf{7} 673--679.

\bibitem[\protect\citeauthoryear{Opgen-Rhein and Strimmer}{2007}]{OS07a}
\textsc{Opgen-Rhein, R.} and \textsc{Strimmer, K.} (2007).
Accurate ranking of differentially expressed genes by a
distribution-free shrinkage approach.
\textit{Statist. Appl. Genet. Mol. Biol.} \textbf{6} 9.
\MR{2306944}

\bibitem[\protect\citeauthoryear{Pomeroy et~al.}{2002}]{PTG02}
\textsc{Pomeroy, S.~L., Tamayo, P., Gaasenbeek, M., Sturla, L.~M.,
Angelo, M.,
McLaughlin, M.~E., Kim, J. Y.~H., Goumnerova, L.~C., Black, P.~M., Lau, C.,
Allen, J.~C., Zagzag, D., Olson, J.~M., Curran, T., Wetmore, C., Biegel,
J.~A., Poggio, T., Mukherjee, S., Rifkin, R., Califano, A.,
Stolovitzky, G.,
Louis, D.~N., Mesirov, J.~P., Lander, E.~S.} and \textsc{Golub, T.~R.} (2002).
Prediction of central nervous system embryonal tumour outcome based
on gene expression.
\textit{Nature} \textbf{415} 436--442.

\bibitem[\protect\citeauthoryear{Sch\"{a}fer and Strimmer}{2005}]{SS05c}
\textsc{Sch\"{a}fer, J.} and \textsc{Strimmer, K.} (2005).
A shrinkage approach to large-scale covariance matrix estimation and
implications for functional genomics.
\textit{Statist. Appl. Genet. Mol. Biol.} \textbf{4} 32.
\MR{2183942}

\bibitem[\protect\citeauthoryear{Schwender, Ickstadt and Rahnenf\"
{u}hrer}{2008}]{SIR08}
\textsc{Schwender, H., Ickstadt, K.} and \textsc{Rahnenf\"{u}hrer,
J.} (2008).
Classification with high-dimensional genetic data: Assigning patients
and genetic features to known classes.
\textit{Biometr. J.} \textbf{50} 911--926.

\bibitem[\protect\citeauthoryear{Singh et~al.}{2002}]{SF02}
\textsc{Singh, D., Febbo, P.~G., Ross, K., Jackson, D.~G., Manola, J.,
Ladd, C.,
Tamayo, P., Renshaw, A.~A., D'Amico, A.~V., Richie, J.~P., Lander, E.~S.,
Loda, M., Kantoff, P.~W., Golub, T.~R.} and \textsc{Sellers, W.~R.} (2002).
Gene expression correlates of clinical prostate cancer behavior.
\textit{Cancer Cell} \textbf{1} 203--209.

\bibitem[\protect\citeauthoryear{Slawski, Daumer and
Boulesteix}{2008}]{SDB08}
\textsc{Slawski, M., Daumer, M.} and \textsc{Boulesteix, A.-L.} (2008).
{CMA}---a comprehensive {Bioconductor} package for supervised
classification with high dimensional data.
\textit{BMC Bionformatics} \textbf{9} 439.

\bibitem[\protect\citeauthoryear{Strimmer}{2008}]{Str08c}
\textsc{Strimmer, K.} (2008).
A unified approach to false discovery rate estimation.
\textit{BMC Bioinformatics} \textbf{9} 303.

\bibitem[\protect\citeauthoryear{Tibshirani et~al.}{2002}]{THNC02}
\textsc{Tibshirani, R., Hastie, T., Narasimhan, B.} and \textsc{Chu,
G.} (2002).
Diagnosis of multiple cancer type by shrunken centroids of gene
expression.
\textit{Proc. Natl. Acad. Sci. USA} \textbf{99} 6567--6572.

\bibitem[\protect\citeauthoryear{Tibshirani et~al.}{2003}]{THNC03}
\textsc{Tibshirani, R., Hastie, T., Narsimhan, B.} and \textsc{Chu,
G.} (2003).
Class prediction by nearest shrunken centroids, with applications to
{DNA} microarrays.
\textit{Statist. Sci.} \textbf{18} 104--117.
\MR{1997067}

\bibitem[\protect\citeauthoryear{Wilson and Hilferty}{1931}]{WH31}
\textsc{Wilson, E.} and \textsc{Hilferty, M.} (1931).
The distribution of chi-square.
\textit{Proc. Natl. Acad. Sci.} \textbf{17} 684--688.

\bibitem[\protect\citeauthoryear{Witten and Tibshirani}{2009}]{WT09}
\textsc{Witten, D.~M.} and \textsc{Tibshirani, R.} (2009).
Covariance-regularized regression and classification for
high-dimensional problems.
\textit{J. Roy. Statist. Soc. Ser. B} \textbf{71} 615--636.

\bibitem[\protect\citeauthoryear{Xu, Brock and Parrish}{2009}]{XBP09}
\textsc{Xu, P., Brock, G.~N.} and \textsc{Parrish, R.~S.} (2009).
Modified linear discriminant analysis approaches for classification
of high-dimensional micoarray data.
\textit{Comput. Stat. Data Anal.} \textbf{53} 1674--1687.

\bibitem[\protect\citeauthoryear{Zuber and Strimmer}{2009}]{ZS09}
\textsc{Zuber, V.} and \textsc{Strimmer, K.} (2009).
Gene ranking and biomarker discovery under correlation.
\textit{Bioinformatics} \textbf{25} 2700--2707.

\end{thebibliography}
\end{document}